# Hybrid Multi Micropattern Gaseous Photomultiplier for detection of liquid-xenon scintillation

Samuel Duval,[a*] Lior Arazi,[b] Amos Breskin,[b] Ranny Budnik,[b] Wan-Ting Chen,[a] Hervé Carduner,[a] A.E.C. Coimbra,[c] Marco Cortesi,[b] Roy Kaner,[b] Jean-Pierre Cussonneau,[a] Jérôme Donnard,[a] Jacob Lamblin,[a] Olivier Lemaire,[a] Patrick Le Ray,[a] J. A. M. Lopes,[c] Abdul-Fattah Mohamad Hadi,[a] Eric Morteau,[a] Tugdual Oger,[a] J.M.F. dos Santos,[c] Luca Scotto Lavina,[a] Jean-Sébastien Stutzmann,[a] and Dominique Thers,[a]

[a]*Subatech, Ecole des Mines, CNRS/IN2P3 and Université de Nantes, 44307 Nantes, France*

[b]*Departement of Astrophysics and Particle Physics, Weizmann Institute of Science, 76100 Rehovot, Israel*

[c]*Instrumentation Centre, University of Coimbra, 3004-516 Coimbra, Portugal*

**Abstract**

Gaseous PhotoMultipliers (GPM) are a very promising alternative of vacuum PMTs especially for large-size noble-liquid detectors in the field of Functional Nuclear Medical Imaging and Direct Dark Matter Detection. We present recent characterization results of a Hybrid-GPM made of three Micropattern Gaseous Structures; a Thick Gaseous Electron Multiplier (THGEM), a Parallel Ionization Multiplier (PIM) and a MICROMesh GAseous Structure (MICROMEGAS), operating in Ne/CF$_4$ (90:10). Gain values close to $10^7$ were recorded in this mixture, with 5.9keV x-rays at 1100 mbar, both at room temperature and at that of liquid xenon (T = 171K). The results are discussed in term of scintillation detection. While the present multiplier was investigated without photocathode, complementary results of photoextraction from CsI UV-photocathodes are presented in Ne/CH$_4$ (95:5) and CH$_4$ in cryogenic conditions.

[*] Corresponding author. Tel +332-51 85 86 49; fax +0-000-000-0000; email: duval@subatech.in2p3.fr

## 1. Introduction

Since the beginning of the 2000s attempts have been made in the field of UV-photon detection with gas-avalanche detectors operating at cryogenic conditions; these aimed at detecting scintillation light in single- and dual-phase liquid argon (LAr) or liquid xenon (LXe) detectors for Functional Medical Nuclear Imaging and Direct Dark Matter Search [1-13]. Matching their operation pressure to that of the noble-liquid detector permits conceiving flat-geometry gaseous photon detectors of large-size, efficient coverage area, and good modularity. Our previous work performed with a double-THGEM/triple-structure device filled with Ne mixture equipped with a reflective CsI photocathode demonstrated the feasibility of detecting LXe scintillation light (178nm) at 173K through a $MgF_2$ window in Ne mixtures [14].

In this article we present recent complementary results of a Hybrid-GPM made of three Micropattern Gaseous Structures: a bare THGEM followed by a PIM and a MICROMEGAS - operating with 5.9keV x-rays under gas-flow of $Ne/CF_4$ (90:10). Gain measurements are compared at liquid xenon (LXe) and room temperatures (T = 171K and 293K, respectively) at a pressure of 1100 mbar. The main objective of combining three different micro-pattern detectors was to reach a high gain and efficient photoextraction with a low avalanche ion-backflow - to minimize photocathode ageing. While the detector was characterized with photocathode, in a parallel study we measured the photoelectron extraction efficiency from CsI at LXe temperature – a key step towards the realization of GPMs for noble-liquid photodetectors.

## 2. Materials and Methods

### 2.1. Experimental set-up

The GPM layout is shown in figure 1 (on the left). It consisted of three successive amplification structures: a THGEM, a PIM, and a MICROMEGAS. The THGEM geometry was: 400μm thickness ("t"), with 300μm hole diameter ("d"), a rim size at the hole edge of 50μm ("r") and a hole spacing of 700μm ("s"). A 70 lines per inch (lpi) nickel cathode grid was placed 10mm above the THGEM. The transfer gap between the THGEM and the PIM was 1.4mm. The PIM consisted of two 5μm thick electroformed nickel grids with different mesh parameters (500 and 670 lpi) isolated from each other by a Kapton spacer defining an amplification gap of 125μm. The spacer was chemically etched into a grid pattern with square openings of 3mm side and 50μm wide insulating separators. The second transfer gap – between 670 lpi nickel grid and MICROMEGAS – was 1.7mm. The MICROMEGAS consisted of a 5μm thick copper electrode with 30μm diameter holes spaced by 60μm. The mesh was maintained at 50μm from a non-segmented anode by Kapton pillars. This multiple structure was enclosed in a stainless-steel cryogenic vessel also shown in figure 1 (on the right).

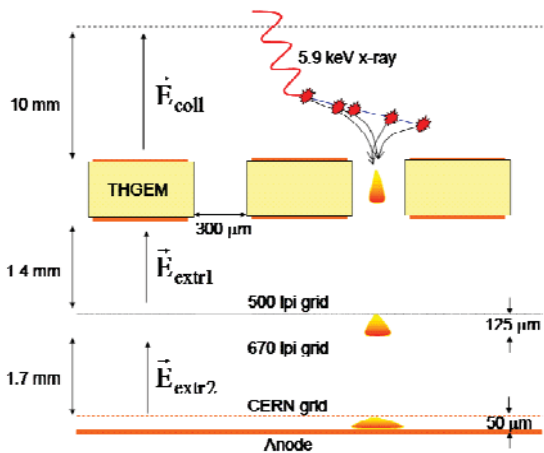 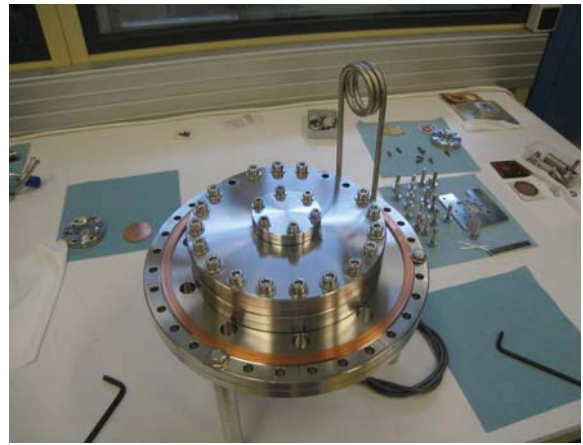

Fig. 1. Schematic layout of the GPM (on the left) and a photograph of the cryogenic GPM (on the right).

The GPM was immersed within ~7 liters of LXe in order to obtain a stabilized gas-mixture temperature of 171K. The cryogenic LXe system and the gas manifold of the GPM are described in [14]. The non-condensable Ne/CF$_4$ (90:10) mixture was flushed through the GPM with gas-flow controllers (EL-FLOW Bronkhorst). A getter (FACILITOR Weld/Purge Gas Purifier Model PS11-FT400-R/N-F) assembled on the GPM gas circuit between the gas flow-meters and the GPM gas inlet was used to purify the Ne mixture before injecting into the GPM chamber. The flow was regulated to 1.5 l/h (at normal conditions). A temperature difference of 2K was monitored by two PT100 temperature sensors located at the gas inlet and outlet. The pressure within the GPM was regulated by the gas-flow controller to 1100 mbar; the gas was evacuated through an oil bubbler to avoid air reflux. The xenon was kept in liquid phase by means of a pulse-tube refrigerator. The gas line, getter system and GPM were pumped for approximately 20 hours with a turbo molecular pump down $7.1 \cdot 10^{-6}$ mbar (measured at the pump entrance), to remove condensable contaminants like water vapor. The pump was stopped and the cryo-system (cryostat/GPM) was sealed during the cooling to LXe temperature. The Ne/CF$_4$ (90:10) gas mixture was then allowed to flow through the getter system to the GPM chamber during one night before applying voltages and performing measurements.

### 2.2. Gain measurements in pulse-mode with a $^{55}$Fe x-rays source

A $^{55}$Fe source was fixed inside the chamber, above the cathode grid (Figure 1). A drift field was applied between the cathode grid and the top of the THGEM to collect primaries induced by x-rays in the conversion gap. The THGEM gain with avalanche-electrons extraction to the next element ($G_{textr}$) was measured by recording the signals on the 500 and 670 lpi grids. The THGEM/PIM/MICROMEGAS gain ($G_{tpm}$) was measured by recording the charge induced pulses on the anode with the preamplifier. Charges were multiplied in the THGEM holes, transferred to the PIM, amplified in the PIM, transferred and finally amplified in the MICROMEGAS.

Pulses were recorded with an ORTEC 142 charge preamplifier and further processed by standard electronics chain; the latter was calibrated by injecting a known-amplitude charge-pulse into the preamplifier input test capacitance.

For gains higher than $10^6$ the anode signal could be recorded directly on a 50Ohms resistance, as shown in Figure 2. The signal was then integrated and differentiated with a fast-filter amplifier (ORTEC 579), was reversed and fed into a peak-sensing ADC Mod V785N. A known input test capacitance permitted the injection for the chain calibration.

### 2.3. Measurements of photoelectron extraction efficiency from CsI at LXe temperature

The photoelectron extraction efficiency from CsI irradiated by UV photons at LXe temperature was studied in a modified version of the setup described in [15]. It comprised a chamber operating either in high vacuum or under flow of Ne/CH$_4$ (95:5) or pure CH$_4$ (experiments involving Ne/CF$_4$ are underway). Measurements at LXe temperature were done by immersing the chamber in ethanol cooled with liquid nitrogen to ~170-180K. The chamber incorporated electrodes and other parts made of stainless steel and ceramics, to minimize outgassing (e.g. water vapor) that could affect the CsI surface, thus the emission, as observed and discussed in [15]. Photocurrent extraction from CsI, deposited on an Al-evaporated polished stainless-steel, was measured by its UV irradiation with an Ar(Hg) lamp (useful wavelength 190 nm), collecting the photoelectrons to a parallel mesh, 6 mm away. The resulting DC current (on the scale of < 3nA) was measured with a Keithley 610C electrometer connected to the photocathode. Measurements were taken at a fixed geometry under the following conditions: (1) vacuum at room temperature; (2) 1050 mbar Ne/$CH_4$ (95:5) and $CH_4$ at room temperature; (3) 1700 mbar Ne/$CH_4$ (95:5) and $CH_4$ at room temperature; (4) 1050 mbar Ne/$CH_4$ (95:5) at 179K and $CH_4$ at 170K. The choice of 1700 mbar in measurement (3) was made to have a reference for the 1050 mbar measurement at 179K, where the average gas density is 1.7 times higher compared to 1050 mbar at room temperature.

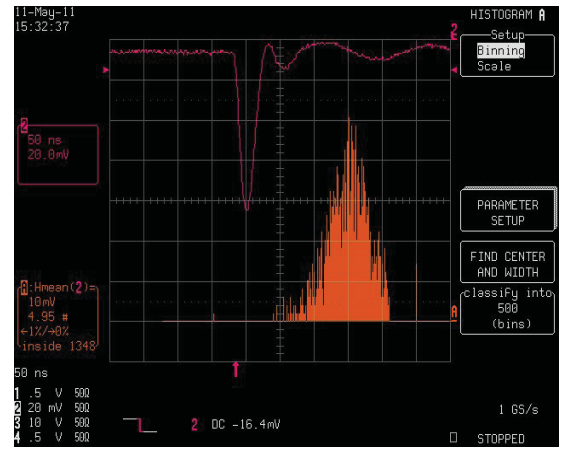

Fig. 2. Oscilloscope snapshot of the $^{55}$Fe charge distribution spectrum (channel A) recorded at LXe temperature (173K) in Ne/CF$_4$ (90:10). A typical pulse is shown in channel 2 (50ns/div; 20mV/div) read on 50Ohms resistance.

### 3. Results

#### 3.1. THGEM gain measurements

Figure 3 shows the single-THGEM gain curve recorded with the $^{55}$Fe source as function of the differential potential ($\Delta V_{THGEM}$) applied, in Ne/$CF_4$ (90:10). The collection field ($E_{col}$) and the extraction field ($E_{extr1}$) (see Fig. 1) were fixed respectively to 0.25kV/cm and 1kV/cm. The last point of the gain curve corresponds to the onset of discharges. Occasional sparks were noticed during THGEM gain measurements (above $\Delta V_{THGEM}$ = 1200V). These occasional sparks, however, did not cause detector instabilities during measurements.

Error bars are associated to the energy resolution (RMS) of the x-rays spectra. One can notice the degradation of energy resolution for $\Delta V_{THGEM}$ values below 1175V, due to the methodology of the measurements. THGEM gain between $\Delta V_{THGEM}$ = 1175V and 1275V was measured by collecting the charges on the 500 lpi PIM grid. Between $\Delta V_{THGEM}$ = 1000V and 1175V the gain was measured by recording the signals on the 670 lpi PIM grid. The rest of the curve (from $\Delta V_{THGEM}$ = 450V to 1000V) was recorded by measuring the gain on the anode. Indeed, successive microstructures were used to record small initial amount of charges extracted from the THGEM (around 200 electrons when gain was closed to one).

#### 3.2. Hybrid-GPM gain measurements

Figure 4 shows THGEM/PIM/MICROMEGAS gain measurements recorded with the $^{55}$Fe x-rays at room- and LXe-temperature in Ne/$CF_4$ (90:10) as function of the MICROMEGAS amplification electric field ($E_{micro}$) for different $\Delta V_{THGEM}$ values. The last point on the LXe-temperature gain curves correspond to the onset of occasional discharges. For

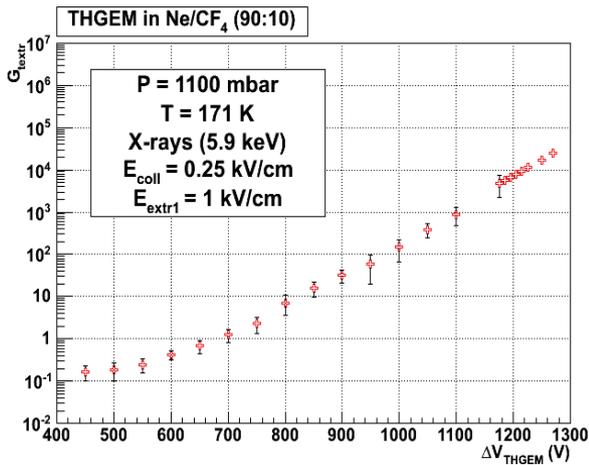

Fig. 3. THGEM gain ($G_{textr}$) in Ne/$CF_4$ (90:10) at 171K as function of the THGEM differential potential ($\Delta V_{THGEM}$) with x-rays at a fixed extraction field ($E_{extr1}$) of 1kV/cm.

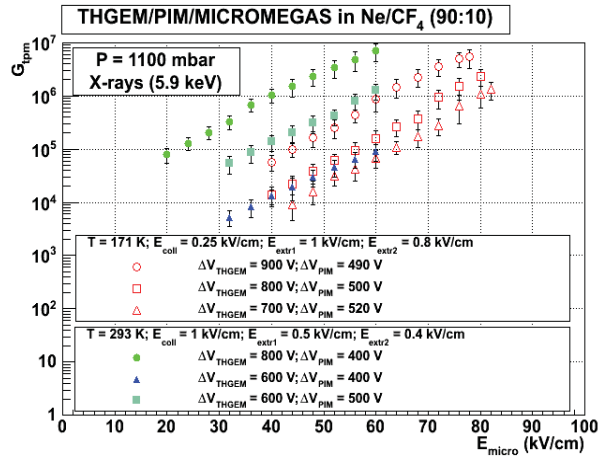

Fig. 4. THGEM/PIM/MICROMEGAS gain ($G_{tpm}$) recorded with x-rays in Ne/$CF_4$ (90:10) at 171 K and 293 K in Ne/CF4.

$\Delta V_{THGEM}$ = 900V the maximum gain recorded at LXe temperature was $5.3 \cdot 10^6$. An example of an energy spectrum is shown in figure 5. Even when the THGEM potential difference was reduced to 800V and 700V (low $G_{textr}$), gains higher than $10^6$ were reached.

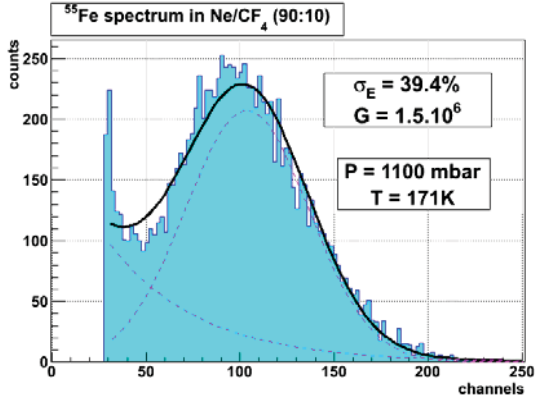

Fig. 5. THGEM/PIM/MICROMEGAS $^{55}$Fe spectrum recorded at 171 K with a 50 Ω resistance instead of the charge-sensitive preamplifier. ($E_{coll}$ = 0.25kV/cm; $\Delta V_{THGEM}$ = 900V; $E_{extr1}$ = 1kV/cm; $\Delta V_{PIM}$ = 490V; $E_{extr2}$ = 0.8kV/cm; $\Delta V_{micro}$ = 330V)

### 3.3. Photoelectron extraction efficiency from CsI at LXe temperature

Figure 6 shows the results of the photocurrent extraction measurements. Figure 6A shows the results obtained in CH$_4$ at room temperature (1050 and 1700 mbar) and at 170K (1050 mbar). The data presented is the photocurrent extraction efficiency, which is defined as the photocurrent at a given pressure and temperature divided by its corresponding value in vacuum for the same applied field. The very high extraction efficiency (>95% at fields larger than ~1.3 kV/cm in all cases) confirms that in CH$_4$ there is relatively little backscattering of extracted photoelectrons back to the photocathode [16]. The extraction efficiency at 1050 mbar is lower at LXe temperature compared to room temperature in fields <1 kV/cm, but becomes approximately the same at higher fields. The extraction efficiency at 1050 mbar in LXe temperature and at 1700 mbar in room temperature are very similar, indicating that the lower efficiency at low field values results practically entirely from residual backscattering of photoelectrons back to the photocathode (with the increased backscattering driven by the 1.7 times higher gas density). Figure 6B shows the photocurrent extraction efficiency for Ne/CH$_4$ (95:5); the effect of backscattering is much larger than in pure CH$_4$ (extraction efficiency <60% at atmospheric pressure at room temperature for fields <1 kV/cm). The extraction efficiency decreases with increased gas density (at 1700 mbar compared to 1050 mbar), and further decreases at 1050 mbar at LXe temperature. The latter observation suggests that the reduction in extraction efficiency at LXe temperature results from additional effects other than increased gas density, possibly due to residual water vapor condensation on the photocathode. For comparison, figure 6B also shows the result of a Monte Carlo simulation for 1050 mbar Ne/CH$_4$ (95:5) at room temperature by Escada *et al.* [17].

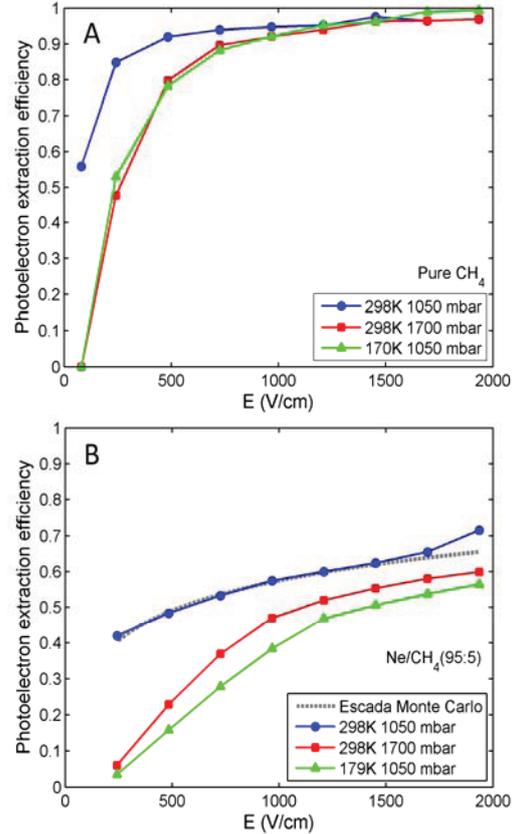

Fig. 6. Effect of temperature and pressure on CsI photocurrent extraction efficiency as a function of the extraction field. (A) Measurements in pure CH$_4$. (B) Measurements in Ne/CH$_4$ (95:5), including the results of a Monte Carlo simulation (Escada *et al* [22]).

## 4. Discussion & Conclusion

### 4.1. Gain measurements

High gains were reached at room- and LXe-temperature, around $10^6$, with the triple-structure detector in Ne/CF$_4$ (90:10), even with low THGEM gain and without sparking induction into the microgaps of the PIM or MICROMEGAS. Such high gains were already observed with triple-GEM detectors in CF$_4$ at room temperature [18]. However, as far as we know no equivalent gains were recorded at liquid xenon temperature. This permits to envisage a simple electronic read-out, in view of lowering the cost of a large-size segmented GPM devices dedicated to local triggering [19].

### 4.2. Ion-backflow

Due to their mass the majority of the ions tends to follow the electric field lines and to flow back to the top electrode in single THGEM devices [20], *i.e.* to the reflective photocathode. This is known to be the origin of photocathode ageing [21]. This effect can be reduced by limiting the number of ions created within the THGEM holes by reducing its gain. In this study the gain of a single-THGEM in Ne/CF$_4$ (90:10) was measured at LXe temperature (171 K). It corroborates the fact that the effective THGEM gain is lower than $10^2$ when the THGEM potential difference is lower than 900V. Future measurements will evaluate the ion-backflow in such conditions with UV light.

### 4.3. Extraction efficiency

The potential difference of 900V corresponds to a surface electric field higher than 1kV/cm for the THGEM geometry involved in this study. Furthermore the amount of gas quencher (CF$_4$) is relatively high. These two parameters should permit to reach good photoelectron extraction efficiency from CsI in the scintillation detection configuration (see reference [14] for details). Indeed, results obtained at room temperature showed that extraction efficiency in Ne/CF$_4$ (90:10) is very close to the one measured in pure CF$_4$ [22]. The results obtained so far with CH$_4$ and Ne/CH$_4$ (95:5) confirms that the primary effect of cryogenic temperatures on the photoelectron extraction efficiency is due to backscattering. Other effects, such as water vapor condensation on the CsI photocathode (observed for Ne/CH$_4$ but not for CH$_4$) are apparently of lower importance and can be limited by effective purification of the gas. Complementary extraction efficiency measurements at LXe temperature in Ne/CF$_4$ are underway aiming to refine our objectives: high extraction efficiency and minimum amplification inside THGEM holes.

**Acknowledgments**

This work was partly supported by the region of Pays de la Loire. Support is acknowledged from FCT Projects CERN/FP/116394/10 and PTDC/FIS/100474/2008, Israel Science Foundation grant 477/10, Minerva Foundation with funding from the German Ministry for Education and Research, and from the Benozyio Foundation. A. Breskin is the W.P. Reuther Professor of Research in The Peaceful Use of Atomic Energy.